\documentclass[prb,aps,twocolumn,showpacs,
amsmath,amssymb]{revtex4}
\usepackage{epsfig}
\begin{document}
\title{Decoherence and relaxation in the interacting quantum dot
  system} 
\author{M. Q. Weng}
\email{weng@ustc.edu.cn.}
\affiliation{
Department of Physics 
University of Science and
Technology of China, Hefei, Anhui, 230026, China}
\date{\today}
\begin{abstract}
In this paper we study the low temperature kinetics of the electrons
in the system composed of a quantum dot connected to two leads by
solving the equation of motion. The decoherence and the relaxation
of the system caused by the gate voltage noise and the electron-phonon
scattering are investigated. In order to take account of the strong
correlation of the electrons in this system, the quasi-exact wave
functions are calculated using an improved matrix product states
algorithm. This algorithm enables us to calculate the wave functions
of the ground state and the low lying excited states with satisfied
accuracy and thus enables us to study the kinetics of the system more
effectively.  
It is found that although both of these two mechanisms are
proportional to the electron number operator in the dot, the kinetics
are quite different. The noise induced decoherence is much more
effective than the energy relaxation, while the energy relaxation and
decoherence time are of the same order for the electron-phonon
scattering. Moreover, the noise induced decoherence increases with the
lowering of the dot level, but the relaxation and decoherence
due to the electron-phonon scattering decrease.  
\end{abstract} 

\pacs{71.27.+a,72.15.Qm,73.21.La}
\maketitle

\section{Introduction} 
\label{sec:intr}

In the past two decades there have been many experimental and theoretical
investigations on the artificial strong interacting quantum system,
especially after the observation of Kondo effect in the
quasi one-dimensional (1D) system composed by a 
semiconductor quantum dot (QD) connected to two leads. 
The Kondo effect of this system produces a perfect transparency with
unitary conductance in the symmetric configuration $\varepsilon_0=-U/2$,
with $\varepsilon_0$ and $U$ being the QD level and on-site Coulomb
repulsion. \cite{lee_88,glazman_88,cronenwett_98,gordon_98,sasaki_00}
Most of the studies focus on the static transport properties. In 
recent years, the kinetics and time evolution of these systems also
attracted much
attention,\cite{nordlander_99,lopz_00,karrai_04,smith_05} as the
study of the kinetics and time evolution enables one to attack the
strong correlated system 
from different perspectives. Different methods have been
employed to study the response of the systems to the ac-modulated or
step switching bias and/or gate 
voltage.\cite{nordlander_99,plihal_00,plihal_05,cazalilla_02} 
Theoretically, the temporal evolution of Kondo-like systems
has been studied by using noncrossing
approximation,\cite{nordlander_99,plihal_00,plihal_05} 
quantum Monte Carlo simulation,\cite{silver_90} 
time-dependent numerical renormalization group,\cite{anders_06}
and time-dependent density-matrix renormalization group (DMRG)
methods.\cite{cazalilla_02} However, only few
studies have considered real dissipation mechanisms, such as
noise and the electron-acoustic (AC) phonon interaction which dominate the
decoherence and relaxation at low temperature.\cite{konig_96} 

In this paper, we study the time evolution of the QD system
under the influence of the gate noise and acoustic phonon by solving
the equation of motion. Since the electrons in this kind of system are
mutually correlated, an improved version of the matrix product states
(MPS) algorithm\cite{ostlund_95,verstraete_04,%
verstraete_05,orensen_05,hand_06,ripol_06}
is employed to study the electron system in order to keep the
many-body effect of the system accurate. The paper is organized as
following. In the second section, we present the system Hamiltonian
and the equation of motion for the time evolution of the system in the
presence of the gate noise and electron phonon interaction. In
the third section we give a brief introduction to the MPS algorithm
and discuss how it can be further improved to study the kinetics of
the QD 
system. We show the numerical results of the decoherence and
relaxation in  the fourth section and summarize in the last section.

\section{System Model}
\label{sec:model}

The system we study is composed of a semiconductor QD connected to two
leads. The Hamiltonian is written as
\begin{equation}
  \label{eq:HamTot}
  H=H_e+H_{ph}+H_{ep}.
\end{equation}
Here $H_e$ and $H_{ph}$ are the Hamiltonian of electron and phonon,
$H_{ep}$ describes the electron-phonon interaction. The electron 
Hamiltonian of a $N$-site QD system is
\begin{equation}
  \label{eq:Ham}
  H_e=\sum_{i\sigma}
  \varepsilon_{i}n_{i\sigma}+Un_{0\uparrow}n_{0\downarrow}
  +\sum_{i\sigma}t_iC^{\dag}_{i\sigma}C_{i+1\sigma}+h.c.\;,
\end{equation}
with $C^{\dag}_{i\sigma} (C_{i\sigma})$ being the creation
(annihilation) operator of electron with spin 
$\sigma(=\uparrow\downarrow)$ at the site $i$. 
The sites with index $i=1,2,\cdots, [N/2] (-[N/2],\cdots,-2,-1)$ are
the sites on right (left) lead, while site with $i=0$ is for the QD. 
In this paper, the on-site energy of leads 
is set to be zero and $t_{i}\equiv t$ for $i=\cdots,-3,-2,1,2,\cdots$. $U$ 
describes the Coulomb repulsion in the QD. The QD level $\varepsilon_{0}$ 
can be controlled by gate voltage and is subjected to the influence of
the noise. For the phonon related part, we only consider the electron
interaction with background longitudinal AC phonon via deformation 
potential. The phonon Hamiltonian and the electron-phonon
scattering are 
\begin{equation}
  \label{eq:H_ph}
  H_{ph}=\sum_{\mathbf{q}\lambda} \omega_{\mathbf{q}}
  b^{\dag}_{\mathbf{q}}b_{\mathbf{q}}
\end{equation}
and 
\begin{equation}
  \label{eq:H_ep}
  H_{ep}=\sum_{\mathbf{q}}\sum_{\sigma}n_{0\sigma}
  M_{\mathbf{q}}(b_{q}+b_{-q}^{\dag})\; ,
\end{equation}
respectively. $b^{\dag}_{\mathbf{q}} (b_{\mathbf{q}})$ is the creation
(annihilation) operator of longitudinal AC phonon with wave-vector
$\mathbf{q}$, whose frequency is $\omega_{\mathbf{q}}=v_{sl}q$ with
$v_{sl}$ standing for the longitudinal sound
velocity. $M_{\mathbf{q}}\propto \sqrt{q}
\langle\Psi_0|e^{i\mathbf{q}\cdot\mathbf{r}}|\Psi_0\rangle$ 
is the corresponding coupling matrix. 
Here $\Psi_0$ is the electron wave-function in the dot. It is chosen
to have Gaussian form $\exp(-r^2/2a^2)$ for simplicity, with $a$
denoting the diameter of the QD.  

The kinetics of 
the system is studied by the temporal evolution of the density
matrix $\rho(\tau)$, whose diagonal elements $\rho_{n,n}(\tau)$
and off-diagonal ones $\rho_{n,m}(\tau)$ stand
for the population of $|n\rangle$ state 
and the coherence between $|n\rangle$ and $|m\rangle$ states at time
$\tau$ respectively. Here $|n\rangle$ and $|m\rangle$ are the
eigenstates of electron Hamiltonian free from the gate noise and
electron-phonon interaction. Using Markovian approximation and secullar
approximation, one can write down the equation of the motion for
the density matrix:\cite{openquantum} 
\begin{eqnarray}
  \label{eq:eom}
&&  {  \partial \rho(\tau)\over \partial \tau}=
-i[H_e, \rho] -i [H_{LS},\rho(\tau)]
-\sum_{\varepsilon}
\Gamma(\varepsilon)
\bigl[
n_0(\varepsilon)
\nonumber\\ && 
n^{\dag}_0(\varepsilon)\rho(\tau)
+\rho(\tau) n_0(\varepsilon)n^{\dag}_0(\varepsilon)
-2 n^{\dag}_0(\varepsilon)\rho(\tau) n_0(\varepsilon)
\bigr]\; .
\end{eqnarray}
Here $H_{LS}=\sum_{\varepsilon} \Delta(\varepsilon)
n_0(\varepsilon)n^{\dag}_0(\varepsilon)$ is the energy shifting due to
electron-phonon interaction. 
$\Delta(\varepsilon)$ and $\Gamma(\varepsilon)$  
are the real and imaginary parts of the following formula
\begin{eqnarray}
 \Delta(\varepsilon)+i\Gamma(\varepsilon)=
 \sum_{\mathbf{q}}|M_{\mathbf{q}}|^2
 \int_0^{\infty} d\tau
 e^{i\varepsilon\tau}
 D(\mathbf{q},\tau)\;,
  \label{eq:DG}
\end{eqnarray}
with $D(\mathbf{q},\tau)$ being the phonon Green function.
$n_0(\varepsilon)=\sum_{n,m}
\delta_{\varepsilon, E_n-E_m}|n\rangle\langle n|n_0|m\rangle\langle
m|$. For the strong electron-optical phonon
interaction, $H_{LS}$ plays important role in the system
properties.\cite{zhu} However, for the weak electron-AC phonon
interaction we consider in this paper, it only slightly renormalizes
the QD level and the 
Counlomb interaction, which is hard to detect experimentally. Therefore
this shifting is simply omitted in this paper. 
For electron-AC phonon interaction via deformation potential 
$\Gamma(\varepsilon)$ takes the following form
\begin{equation}
  \label{eq:Gamma}
  \Gamma(\varepsilon)=
  \Delta_p
  |\varepsilon|^3 [(1+n_B(\varepsilon))\theta(\varepsilon)
  +n_B(-\varepsilon)\theta(-\varepsilon)]\,,
\end{equation}
where $n_B(\varepsilon)=1/(e^{\varepsilon/T}-1)$ is the Bose function
at temperature $T$, and $\theta(\varepsilon)$ is the 
Heaviside step function. 
$\Delta_p$ depends on the material and structure dependent parameter
as well as on the electron-phonon coupling. 

\section{Numerical Scheme}
\label{sec:mps}

To solve the equation of motion, one needs the the wave-function of
eigenstates of the electron system to obtain the matrix elements of
density matrix and $n_0$. Since the electrons governed by the
Hamiltonian [Eq. (\ref{eq:Ham})] form a mutually correlatted system,
it is important to take the correlation into account in the study of
the kinetics at low temperature. 
Here we use an improved MPS algorithm to obtain the many-body
wave-functions so that the strong correlation of the electron system
can be automatically taken into account.  

For 1D ground state problem, MPS is known to be equivalent to
DMRG.\cite{ostlund_95,verstraete_04,ripol_06}
DMRG was first proposed to study the ground and low-lying excited
states of quantum systems,\cite{white_92,white_93,dmrg_book} and later
was further extended to the simulation of the time evolution, 
calculation of excitation spectra and finite temperature properties of
quantum systems.\cite{white_04,vidal_03,ripol_06} 
The success of DMRG is eventually understood by its connection to  
MPS.\cite{ostlund_95,verstraete_04,ripol_06}
In the MPS algorithm, the wave-function of a $N$-site lattice is represented
by a group of $A$-matrices whose dimension is usually much smaller than
the dimension of Hilbert space
\begin{eqnarray}
  \label{eq:MPS}
  \sum\limits_{s_1\cdots s_{i-1} 
    \atop s_is_{i+1}\cdots s_L}&&
  \mbox{Tr}\bigl\{A^{s_1}\cdots A^{s_{i-1}}
  A^{s_i}A^{s_{i+1}}\cdots A^{s_L}\bigr\}\nonumber \\ 
  &&  |s_1\cdots s_{i-1}s_{i}s_{i+1}\cdots s_L\rangle\ , 
\end{eqnarray}
with $s_i$ representing the local state index at $i$-th site. To start
the calculation, an 
initial wave-function is given by other calculations or randomly
generated. The wave-function is then gradually optimized by a process
called ``sweep'': The wave-function is optimized 
by only minimizing the energy with regard to 
the local $A$-matrix at one or two ``center'' sites 
while keeping other $A$-matrices unchanged. 
One then moves the ``center'' site to the left or right neighbor 
by targeting to the ground state\cite{white_92,white_93}
or performing singular value decomposition.\cite{verstraete_04} 
The process repeats until a desirable accuracy is
achieved or the accuracy can not be further improved by more sweeps.
In traditional DMRG algorithm, if more than 
one eigen-state are to be calculated, they should be targeted simultaneously
during the sweep process. This requires larger $A$-matrix dimension
and more computing resources. The more eigen-states are required,
the less accurate they are for a fixed $A$-matrix
dimension. Therefore, the traditional DMRG algorithm is usually
limited to the calculation of the ground state and few low lying
excited states. Here we propose an algorithm to improve the
calculation of the excited states by using MPS. 

\begin{figure}[htbp]
  \centering
  \epsfig{file=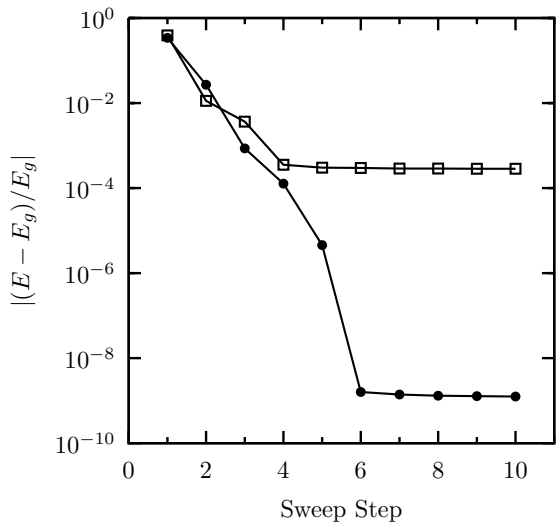,width=0.85\columnwidth}
  \caption{Relative error of the ground state energy vs. sweep
    step for the system with $N=17$, and
    $N_{\uparrow}=N_{\downarrow}=8$. The hopping between the leads and
    the QD is $0.1t$ and the on-site interaction $U=6t$.  
    The filled dots are the results of
    improved algorithm; The open boxes are those of single sweep with 
    randomly generated initial wave-function. The curves are the
    guilds for the eyes. 
  }
  \label{fig:Ev_Step}
\end{figure}

In our algorithm, instead of calculating all of the required
eigen-states simultaneously, they are calculated step by step. We
first calculate the ground state by using the traditional DMRG method 
and obtain the wave-function in the normal 
MPS form like Eq.~(\ref{eq:MPS}).  
To obtain the first excited state, we again
generate an initial wave-function then gradually optimize it
through sweep. However, during the sweep, we first
orthogonalize the excited state and the ground state by performing
Gramm-Schmidt orthogonalization on the local $A$-matrix at the center
site then minimize the energy for the excited state. In this way, we can
obtain the wave-function of the first excited state in the normal MPS
form when the sweep converges. Since the wave-function of ground state
is in the normal MPS form, the sweep of the first excited state 
does not change the ground state.\cite{cirac_07} 
Similarly, to calculate the higher excited states, we first perform
the orthogonalization of wave-function to all of the lower states then
optimize the energy during the sweep. By repeating this process, we
can get the excited states step by step without sacrificing the
accuracy of the lower states.

In our algorithm, the normal MPS form of wave-functions are required
to carry out the sweep of the excited states. 
There are two major algorithms for sweep: two-site sweep and one-site 
sweep.\cite{white_93,dmrg_book} Two-site sweep gives better
eigen-energy since it uses a larger Hilbert space and therefore it is
more memory hungry and CPU time consuming. Moreover, the wave-function
obtained by two-site sweep is no longer in the normal MPS form.   
For example, if the center sites are $i-1$ and $i$, the corresponding
matrix of the wave-function are 
$\cdots, A^{s_{i-2}}$, $A^{s_{i-1},s_i}$, $A^{s_{i+1}}, \cdots$. One
has to break the matrix $A^{s_{i-1},s_i}$ into the multiple of two
Matrices with larger dimension to get the normal MPS form. 
The wave-function loses its optimality if these two
matrix are truncated to the original dimension. 
On the other hand, one-site sweep is cheaper to carry out. 
More importantly, 
the wave-function of one-site sweep algorithm is always in MPS form,
therefore it keeps the optimality when the center site moves. 
However, one-site sweep can easily fall into some
local optimal points and usually fails to give satisfied eigen-energy
and wave-function unless a good initial wave-function is
given to start the sweep.  
Our solution to the dilemma is to use the combination of two-site and
one-site sweep. That is, we first use two-site sweep to
improve a randomly generated wave-function and truncate the improved 
wave-function to the normal MPS form and use it as the initial 
wave-function for the one-site sweep algorithm. It is expect that,
the truncated wave-function would be a much better initial function
even when the initial two-site sweep is not converged.

In the following we first use this hybrid sweep algorithm to study
the ground state of the QD system to demonstrate its feasibility.
In Fig.~\ref{fig:Ev_Step} the relative error of ground state energy
is plotted as a function of the number of the
sweep. In the calculation, we keep the most relevant 256 states,
and the ``true'' ground state energy $E_g$ is obtained by two-site
sweep with 400 states kept. In the calculation, the
total site number, the spin-up and -down electron number are $N=17$,
and $N_{\uparrow}=N_{\downarrow}=8$ respectively. The hopping between
the QD and the leads is chosen to be $t_{-1}=t_0=0.1t$.
In the QD, the on-site energy
$\varepsilon_0=-2t$ and the Coulomb repulsion $U=6t$. For
comparison, we also plot the result of pure one-site sweep with a
random initial wave function. It is noted that
the first three steps of our new algorithm are two-site sweep. 
One can see from the figure that the initial two-site sweep gives
very good start point for one-site sweep even when the
initial two-site sweep is not converged. The ground state energy
obtained by this algorithm is very close to that obtained by the
two-site sweep. More importantly, with this hybrid sweep
algorithm, one not only gets an accurate eigen-energy but also a very
accurate wave-function with cheaper price. This sweep algorithm also
works for the excited states. For higher excited states, the error
in the eigen-energy is usually larger than those of ground state and
lower excited states. However, when we keep up to 256 states during
the sweep the accuracies of the lowest 10 states are desirable. 

\begin{figure}[htbp]
  \centering
  \epsfig{file=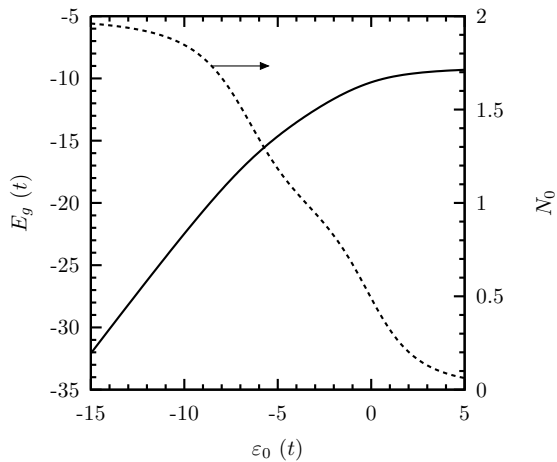,width=0.85\columnwidth}
  \caption{Ground state energy $E_g$ (solid curve) and the occupation
    $N_0$ (dashed curve) of the QD vs. the QD level. Noted that the
    axis for $N_0$ is on the right. The hopping between the leads and
    the QD is $0.1t$ and the on-site interaction $U=6t$. The site and
    the electron numbers are $N=9$, $N_{\uparrow}=N_{\downarrow}=4$
    respectively.  
  }
  \label{fig:Eg}
\end{figure}

\begin{figure}[htbp]
  \centering
  \epsfig{file=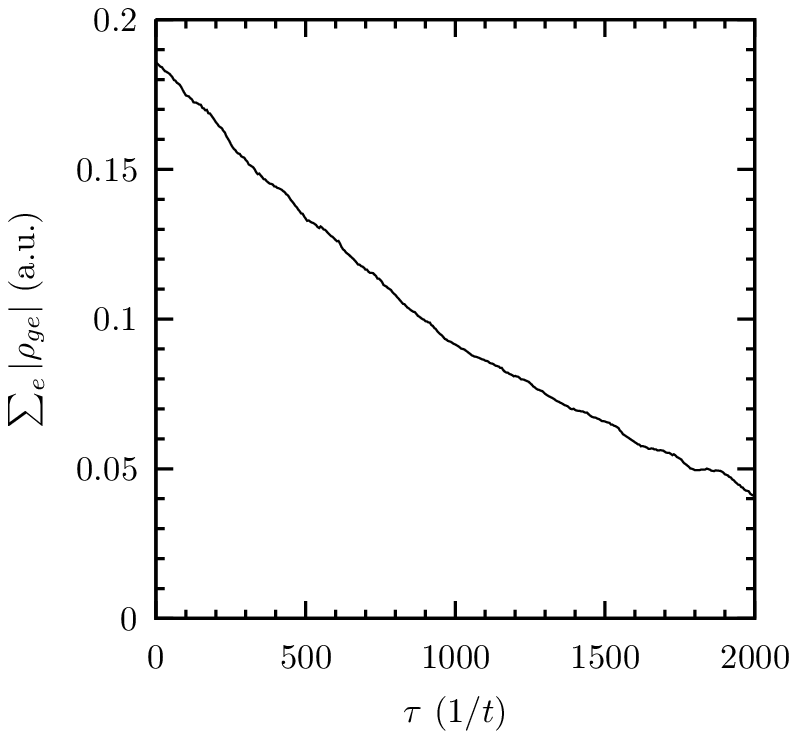,width=0.85\columnwidth}
  \epsfig{file=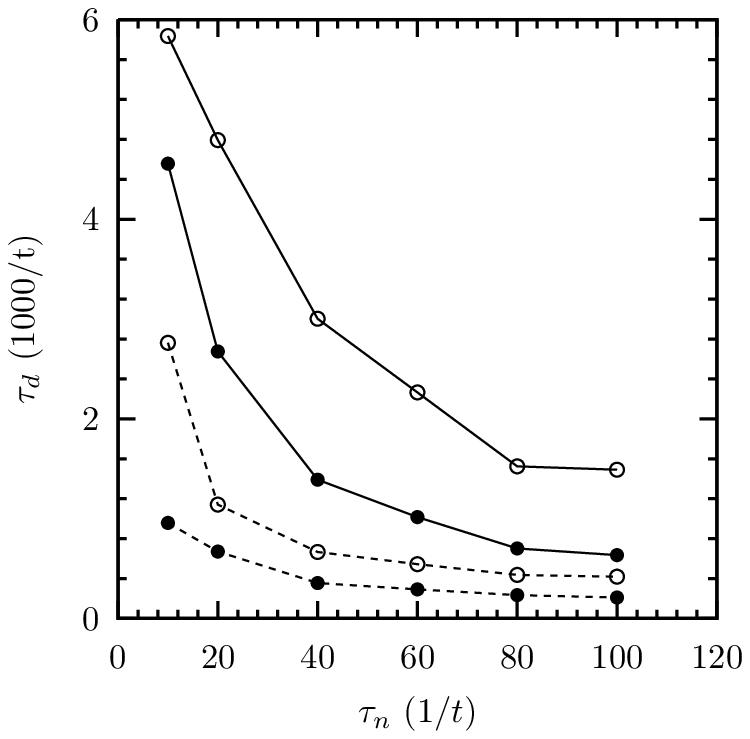,width=0.85\columnwidth}
  \caption{ a) The temporal evolution of the averaged coherence between
    the ground state and the excited states for the QD level
    $\bar{\varepsilon}_0=-3 t$, $\Delta V=0.5 t$ and $\tau_n=40/t$.
    b) The decoherence time as a function of noise characteristic time
    $\tau_n$ for different situations: Solid curves are those for
    $\Delta V=0.5 t$ and dashed ones for $\Delta V=1 t$; Filled circles
    and open circles are for $\bar{\varepsilon}_0=-3$ and $-2 t$
    respectively. 
  }
  \label{fig:noise}
\end{figure}

We now use this method to calculate the wave-functions and
the eigen-energies of the system at different gate voltages. The
ground state energy and the electron occupation number in the QD
of the system with $N=9,N_{\uparrow}=N_{\downarrow}=4$ are plotted
in the Fig.~\ref{fig:Eg}. One can see that for positive on-site
energy, the electron occupation in the QD is small and the ground
state energy is almost independent of gate 
voltage. For large negative QD level, the QD is doubly
occupied and the ground state energy become a linear function
of the QD level. The slope of the function is $2$, which agrees
with the result of $N_0\simeq 2$. 
In the regime of $-6t\le\varepsilon_0\le -2t$, roughly
corresponding to the transparent regime ($-U\le
\varepsilon_0\le 0$), it can be seen that the rate of the changing of
occupation {\it vs.} the changing of the QD level is distinctly slower than
that of its neighbor. The abnormality in this regime may be related
to formation of Kondo singlet which is the superposition of the
localized and the delocalized states. Since the delocalized states have
lower occupation number in the QD, the formation of the Kondo singlet 
slows down the changing rate of the occupation. 

\section{Dephasing and Decoherence}
\label{sec:result}

Once we have the wave-functions, we can simulate the kinetics of the
interacting system under the gate voltage noise and the
electron-phonon interaction. In the low temperature regime, it is
expected that only the lowest few states are involved. Therefore we
can study the kinetics by solving the equation of motion in the
truncated Hilbert space composed of lowest ten states. 
We first study the
decoherence caused by the noise on the QD level by solving the
equation of motion. In the presence of the noise, the QD level becomes 
$\varepsilon_0=\bar{\varepsilon}_0+V(\tau)$, where $\bar{\varepsilon}_0$
is the average QD level, and $V(\tau)$ is the noise due to the 
fluctuation of gate voltage. 
For a system free from noise, the coherence between different states,
{i.e.} the off-diagonal elements of density matrix, evolve as
$\rho_{n,m}(0)e^{-i(E_n-E_m)\tau}$ whose amplitudes do not decay with
time. The noise irregularly shifts the energy levels with different
amounts for different states and causes transition between the
states. Since the amplitude of the noise randomly changes, 
different states eventually lose their phase information over time. As
a result the amplitude of the coherence decays with time. 
In the Fig.~\ref{fig:noise}~(a) we plot the time evolution of the
coherence between the ground state and all of the excited states
$\sum_{e}|\rho_{ge}(\tau)|$ under different conditions. To simulate
the kinetics of suddenly switch of gate voltage, the initial
wave-function wave-function is chosen to be the unperturbed wave
function of zero gate voltage in the truncated space. The density
matrix is averaged over 1000 samplings with the noise $V(\tau)$ evenly
distributed over the regime $[-\Delta V,\Delta V]$ to reflect the
randomness of the noise. One can see from the figure that the
coherence damped oscillates with  
time. The envelope of the temporal evolution curve can be fitted
by an exponential function $\propto e^{-\tau/\tau_d}$ with a
decoherence time $\tau_d$. 
The decoherence times under different
conditions are plotted in the Fig.~\ref{fig:noise}~(b). One can see that
decoherence time decreases with the increase of the noise amplitude and 
the noise characteristic time $\tau_n$. The on-site energy of the QD
also affect 
the decoherence rate, the higher QD level the slower the system loses
its coherence. The boost of the decoherence rate by the increase of
noise level is quite easy to understand. When the amplitude of the
noise increases, the transition rates and the difference of energy
shifts become larger. Therefore different states lose the phase
information more quickly. The decrease of the decoherence rate with
the decrease of $\tau_n$ is due to the suppression of the unitary time
evolution by the irregular change of the noise. This is similar
to the motion narrowing effect.\cite{slichter} From the Hamiltonian,
one can see that the perturbation of the noise is proportional to 
the electron number in the QD. Therefore, the average QD level also
affects the kinetics of the QD system. The lower the QD level, the
larger the electron number in the QD, and the smaller the decoherence
time. It should be noted that although the gate noise is efficient to
remove the phase coherence, it is very inefficient to bring the system
to the thermal equilibrium state as it does not directly carry the
energy away from the electron system. The relaxation to the thermal
equilibrium caused by the noise is more than ten times slower than the
decoherence. 

\begin{figure}[htbp]
  \centering
  \epsfig{file=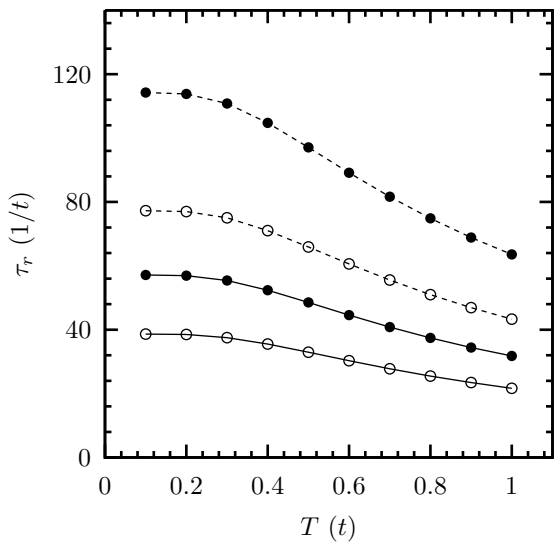,width=0.85\columnwidth}
  \caption{
    The relaxation time (solid curves) and the decoherence
    time (dashed curves) as functions of temperature $T$. Filled and
    open circles are for $\bar{\varepsilon}_0=-3$ and $-2 t$
    respectively. 
  }
  \label{fig:tphonon}
\end{figure}

We now study the relaxation and decoherence due to the 
electron-phonon interaction. In the Fig.~\ref{fig:tphonon} 
we plot the relaxation time $\tau_r$ and the decoherence
time $\tau_d$ as functions of temperature $T$ under different gate
voltages. Noted that $\Delta_p$ is chosen to be $t$ in the
calculation. One can see from the figure that both $\tau_r$ and
$\tau_d$ increases as the temperature decreases and saturate at low
temperature as the electron-phonon scattering saturates to the
emitting phonon limit $\Gamma(\varepsilon)\sim
\Delta_p\varepsilon^3\theta(\varepsilon)$.  There are two qualitative
differences between phonon induced relaxation/decoherence and noise
induced relaxation/decoherence, even through their perturbation
Hamiltonians are both 
proportional to $n_0$, the electron number operator in the QD.
Firstly, the phonon induced relaxation is
a little bit faster than the decoherence, while noise induced
relaxation is more than ten times slower than the decoherence. 
Moreover, the phonon induced relaxation and decoherence decrease with
the lowing of the QD level. The differences rise from the fact that
electron-phonon scattering is inelastic scattering. The electron
system approaches to thermal equilibrium through the absorbing or
emitting of 
phonon. As the electron system loses both energy and phase information
by absorbing/emitting phonon, $\tau_r$ and $\tau_d$ are of the same
order. Furthermore, the main contribution of $n_0$ operator to the
phonon induced relaxation/decoherence comes from the off-diagonal
elements $n_0(\varepsilon)$ with $\varepsilon\not =0$. For lower QD level,
the electrons are more deeply trapped in the QD and it is 
harder for them to hop to
higher-energy states. On the other hand, the main contribution of
$n_0$ to the noise induced decoherence is the diagonal terms, that is
the average electron occupation number in the QD. Therefore the lower
QD level, the quicker the noise induced decoherence. 

\section{Conclusion}
\label{sec:conclusion}

In conclusion, we propose an improved matrix product states 
algorithm to calculate the excited states and 
use this algorithm to study the relaxation and decoherence caused by
the noise and electron-phonon interaction in the 
interacting QD system. Although both of these two mechanisms are
proportional to the electron number operator $n_0$ in the QD, 
the kinetics due to these two mechanisms are quite different. The
noise shifts the energy with different amounts for different states
and causes transition between states. The irregular change of the
noise results in the losing of the phase information of the electron
system and hence of the 
decoherence. However, the noise does not directly carry energy from
the electron system, therefore the energy relaxation due to noise is
very inefficient. When the electron-phonon interaction is present,
the electron system relaxes to the thermal equilibrium by absorbing or
emitting phonon. As the electron system loses both energy and phase
information after the scattering, the energy relaxation and
decoherence time are the same order for electron-phonon scattering. 
Moreover, the main effect of the noise comes from the diagonal terms
of $n_0$, while for electron-phonon scattering it is from the
off-diagonal terms. As a result, the noise induced decoherence
increases with the lowering of the QD level, but the relaxation and
decoherence due to electron-phonon scattering decrease. 

\begin{acknowledgements}
  The author would like to thank M. W. Wu 
  for the inspiring and valuable discussions. 
  This work is supported by Natural Science
  Foundation of China under Grant No. 10804103, the National Basic
  Research Program of China under Grant No.\ 2006CB922005 and the
  Innovation Project of Chinese Academy of Sciences.
\end{acknowledgements}


\end{document}